# Unidirectional perfect reflection and radiation in double-lattice photonic crystals

Takuya Inoue,[1,*] Naoya Noguchi,[1] Masahiro Yoshida,[2] Heungjoon Kim,[2] Takashi Asano,[2] Susumu Noda[1,2*]

[1]Photonics and Electronics Science and Engineering Center, Kyoto University,

[2]Department of Electronic Science and Engineering, Kyoto University,

Kyoto 615-8510, Japan

*To whom correspondence should be addressed.

E-mail: t_inoue@qoe.kuee.kyoto-u.ac.jp, snoda@kuee.kyoto-u.ac.jp.

**Abstract**

**Non-Hermitian photonic systems are known to exhibit unique phenomena, where non-Hermiticity is typically introduced by material loss or gain. Here, we propose and experimentally demonstrate unidirectional phenomena solely based on radiation. Our design is on the basis of a double-lattice photonic crystal that has a linear dispersion with a single exceptional point, where the magnitudes of Hermitian and non-Hermitian couplings are cancelled out in one direction. Based on this concept, we realize a unidirectional waveguide which shows perfect radiation when light is incident from one side, and shows perfect reflection when light is incident from the other side. Our results will open up a new route toward harnessing non-Hermiticity.**

Non-Hermitian photonic systems, in which conservation of energy does not hold, are known to exhibit various unique optical phenomena in the vicinity of exceptional points, where both eigenvalues and their eigenvectors coalesce [1-10]. Unidirectional reflectionless light propagation [11] is one notable example, where the reflectivity of light is asymmetric depending on the direction of incidence. This phenomenon is induced by unidirectional couplings between two counterpropagating modes and resultant coalescence of the two eigenmodes at the exceptional point, which can be realized only in non-Hermitian systems. So far, various unidirectional devices based on the introduction of absorption loss in the materials have been investigated [12-14], where the modulation of the real and imaginary part of the dielectric permittivity causes the asymmetric wave scattering in the waveguide. However, the introduction of the material absorption loss inevitably causes the waveguided light to be absorbed unless gain is introduced, which inhibits the realization of perfect reflection and the extraction of the waveguided light for use in real applications.

In this paper, we propose a unidirectional waveguide solely based on vertical radiation from a photonic crystal, which shows perfect radiation when light is incident from one side and perfect reflection when light is incident from the opposite side as shown in Fig. 1(a). Here, a double-lattice photonic crystal composed of an elliptic hole and a circular hole [15,16] is introduced in a silicon waveguide. The origin of the unidirectionality is the Hermitian and non-Hermitian couplings between the counter-propagating basic waves ($k_x=\pm 2\pi/a$) in Figs. 1(b) and 1(c) [16]. Here, the amplitudes of the basic waves are defined as $R_x$ and $S_x$, respectively, $R$, $I$, and $\mu$ are all real numbers, and the position of the air holes in the unit cell are adjusted so that the phase of the non-Hermitian cross-couplings $\theta_{pc}$ [16] is equal to $\theta_{pc}=0$. Figure 1(b) shows the Hermitian couplings without accompanying energy loss ($R\pm iI$), which is mainly determined by the Fourier coefficients that induce the 2$^{nd}$-order diffraction with a reciprocal vector of $G_{\pm 2}=\pm 4\pi/a$. In the case of double-lattice photonic crystals, the real and imaginary part of the Hermitian couplings can be independently controlled by the hole distance ($d$) and the hole size asymmetry ($S_1/S_2$) of the double-lattice photonic crystal, respectively [15,16]. On the other hand, photonic crystals with 180°-rotational symmetries (e.g. circular-hole photonic crystals [17] or rectangular gratings [18]) always satisfy $I=0$. Figure 1(c) shows the non-Hermitian cross-couplings with radiated loss ($i\mu$), whose magnitude is determined by the Fourier coefficients that induce the 1$^{st}$-order diffraction with a reciprocal vector of $G_{\pm 1}=\pm 2\pi/a$. Using these coefficients, the complex photonic band structure near the $\Gamma$ point (frequency: $\delta$, radiation constant: $\alpha$) can be analyzed as follows in the framework of coupled-wave theory [16,19];

$$(\delta + i\alpha/2)\begin{pmatrix} R_x \\ S_x \end{pmatrix} = \mathbf{C}\begin{pmatrix} R_x \\ S_x \end{pmatrix} + \begin{pmatrix} \Delta k_x R_x \\ -\Delta k_x S_x \end{pmatrix}, \qquad (1)$$

$$\begin{aligned}\mathbf{C} &= \mathbf{C}_{\text{Hermitian}} + \mathbf{C}_{\text{non-Hermitian}} \\ &= \begin{pmatrix} 0 & R+iI \\ R-iI & 0 \end{pmatrix} + \begin{pmatrix} i\mu & i\mu \\ i\mu & i\mu \end{pmatrix} \\ &= \begin{pmatrix} i\mu & R+iI+i\mu \\ R-iI+i\mu & i\mu \end{pmatrix}.\end{aligned} \qquad (2)$$

Here, $\mathbf{C}$ is a coupled-wave matrix, $\Delta k_x$ denotes the wavenumber deviation from the $\Gamma$ point, and $\delta$ denotes the deviation from the frequency of the empty lattice at the second-order $\Gamma$ point. $\mathbf{C}_{\text{Hermitian}}$ is a matrix that represents the Hermitian couplings, where the condition $\mathbf{C}_{\text{Hermitian}} = \mathbf{C}^{\dagger}_{\text{Hermitian}}$ is satisfied, and $\mathbf{C}_{\text{non-Hermitian}}$ is an anti-Hermitian matrix which represents the non-Hermitian couplings, where the condition $\mathbf{C}_{\text{non-Hermitian}} = -\mathbf{C}^{\dagger}_{\text{non-Hermitian}}$ is satisfied. The magnitude of the diagonal terms and non-diagonal terms in $\mathbf{C}_{\text{non-Hermitian}}$ are the same ($\mu$) for the radiative couplings [16, 19], which makes it possible to realize perfect reflection as explained later. It should be noted that the latter becomes smaller than the former for material-loss-induced couplings without gain [12].

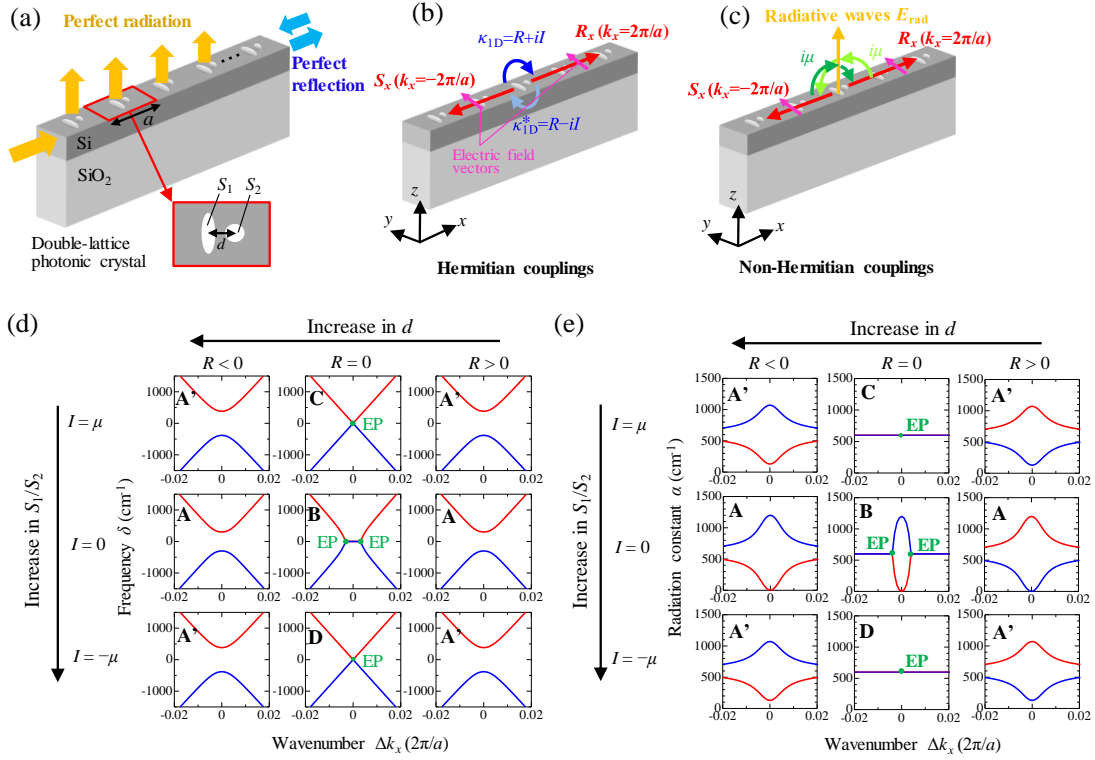

**FIG. 1.** **(a)** Schematic of the proposed unidirectional device with a double-lattice photonic-crystal. **(b)** Hermitian cross couplings between the basic waves. **(c)** non-Hermitian cross couplings between the basic waves. **(d)(e)** Calculated photonic band structures (d) and radiation constants (e) near the Γ-point when the Hermitian coupling coefficients ($R$ and $I$) are varied and the non-Hermitian coupling coefficient is fixed ($\mu$=300 cm-1). Symbols in each graph are as follows; A: $R\neq0$ and $I$=0, A': $R\neq0$ and $I\neq0$, B: $R$=0 and $I$=0, C: $R$=0 and $I$=$\mu$, D: $R$=0 and $I$=$-\mu$. Unidirectional perfect reflection and radiation shown in (a) can be realized at a single exceptional point (EP) in the panels C and D.

Figure 1(d) and 1(e) shows the calculated photonic band structures ($\delta$) and radiation constants ($\alpha$) near the Γ-point when the Hermitian coupling coefficients ($R$ and $I$) are varied and the magnitude of the non-Hermitian coupling coefficient is fixed ($\mu$=300 cm$^{-1}$). When the real part of the Hermitian coupling coefficient ($R$) is not zero (most typical case), a photonic band gap (PBG) opens as shown in the panels A ($I=0$) and A' ($I\neq0$). When both the real part and imaginary part of the Hermitian coupling coefficients ($R$ and $I$) are exactly zero [17,18], the photonic band has two exceptional points (EPs) at non-zero wavenumbers as shown in the panel B. As described in the previous paragraph, photonic crystals with 180°-rotational symmetries always satisfy $I=0$, and thus they exhibit either of the photonic bands shown in the panels A or B. On the other hand, the double-lattice photonic crystal shown in Fig. 1(a) can control the magnitude of $I$ by changing the hole-size asymmetry ($S_1/S_2$), and thus this photonic crystal can realize a linear dispersion with a single EP when $I$ is equal to $\pm\mu$ as shown in the panels C and D. In these special cases, only one of the non-diagonal cross-coupling terms in the coupled matrix **C** in Eq. (2) (namely, $R+iI+i\mu$ or $R-iI+i\mu$) becomes zero, which realizes unidirectional wave scattering between the two basic waves at the frequency of the EP.

More quantitatively, light propagation and vertical radiation inside the finite-sized double-lattice photonic crystal ($0 \leq x \leq L$, where $L$ is the length of the waveguide) at the single EP shown in the panel C ($R=0$, $I=\mu$) can be analyzed with the following finite-sized coupled-wave equation [20];

$$\frac{\partial}{\partial x}\begin{pmatrix} R_x \\ -S_x \end{pmatrix} = i\mathbf{C}\begin{pmatrix} R_x \\ S_x \end{pmatrix} = \begin{pmatrix} -\mu & -2\mu \\ 0 & -\mu \end{pmatrix}\begin{pmatrix} R_x \\ S_x \end{pmatrix}, \qquad (3)$$

$$E_{rad} = \sqrt{2\mu}\left(R_x + S_x\right). \qquad (4)$$

Here, $E_{rad}$ is the electric field of the radiative wave in the vertical direction. By analytically solving Eqs. (3) and (4), we can predict the unidirectional reflection and radiation phenomena in the proposed device. When we consider the boundary condition where $R_x$ is incident from one side ($R_x = E_0$ at $x = 0$) and $L$ is much longer than $1/\mu$, the analytical solutions for $x > 0$ are

$$R_x = E_0 \exp(-\mu x)$$
$$S_x = 0 \qquad (5)$$
$$E_{rad} = \sqrt{2\mu} E_0 \exp(-\mu x)$$

In this case, all of the incident light is radiated in the vertical direction without reflection since $S_x = 0$. On the other hand, when $S_x$ is incident from the other side ($S_x = E_0$ at $x = L$), the solutions for $x < L$ are

$$R_x = -E_0 \exp[\mu(x-L)]$$
$$S_x = E_0 \exp[\mu(x-L)] \qquad (6)$$
$$E_{rad} = 0$$

In this case, all of the incident light is reflected backward without being radiated since $|S_x| = |R_x|$ and $E_{rad} = 0$. Therefore, we can realize perfect radiation in the vertical direction (without reflection) when light is incident from one side, and perfect reflection (without radiation) when light is incident from the opposite side. In the conjugate case (at the single EP shown in the panel D, where $R=0$, $I=-\mu$), the incidence direction in which perfect radiation and perfect reflection occur is reversed. It should be noted that perfect reflection in Eq. (6) is realized only when the magnitudes of the diagonal terms and non-diagonal terms of $\mathbf{C}_{\text{non-Hermitian}}$ in Eq. (2) are balanced, which is one notable feature of the unidirectional waveguides solely based on radiation.

To confirm the above theoretical predictions, we perform numerical simulations of light propagation in double-lattice photonic-crystal waveguides using a three-dimensional finite-difference time-domain (FDTD) method. The lattice constant $a$, the width $w$, the thickness $t$, and the total area of the two air holes ($S_1+S_2$) are fixed to $a$=680 nm, $w$=600 nm, $t$=220 nm, and $S_1+S_2=0.08a^2$, respectively. In the FDTD simulations, we first calculate photonic band structures of the double-lattice photonic crystals using a Bloch boundary condition. Specifically, we investigate the structural parameters of the double-lattice photonic crystal ($d$ and $S_1/S_2$) to realize photonic band structures shown in the panels A-D in Fig. 1(d) (below, we refer to the designed structures as Designs A-D, respectively). Then, we calculate the reflection and radiation spectra of the designed waveguides for light incident from the left and right sides. In these calculations, the light source is placed inside the coupling waveguides without air holes connected to the double-lattice photonic crystals, and the length $L$ of the photonic crystal is assumed to be long enough (320$a$) to enable perfect radiation without transmission. Figure 2 shows the calculated photonic band structures and reflection/radiation spectra for left-incident and right-incident light for Designs A-D. In these figures, the radiation efficiency $e$ is calculated based on the sum of the upward and downward radiation (note that the introduction of vertical asymmetry may also enable the cancellation of the downward radiation [21,22]). Figure 2(a) shows the calculated result for Design A ($R$>0, $I$~0), which has a PBG at the $\Gamma$ point. Here, the reflection (radiation) spectra become almost the same for both incident light owing to the symmetric mutual couplings between the two basic waves in the non-diagonal terms of the coupled-wave matrix **C** ($R+iI+i\mu = R-iI+i\mu$ when $I$=0). Design B ($R$~0, $I$~0), which has two EPs at the off-$\Gamma$ points, also shows no dependence for the incident directions as shown in Fig. 2(b). On the other hand, Design C ($R$~0, $I$~$\mu$) and D ($R$~0, $I$~$-\mu$), which have linear dispersions with single EPs, exhibit totally different properties from Designs A and B; in these structures, we obtain nearly perfect radiation ($e$>0.99) for light incident from one side and nearly perfect reflection ($r$>0.99) for light incident from the other side at the frequency of the EP. In addition, the dependence of the radiation/reflection characteristics on the incident direction is reversed for Design C and Design D. These results clearly validate our theoretical predictions in Eqs. (5) and (6).

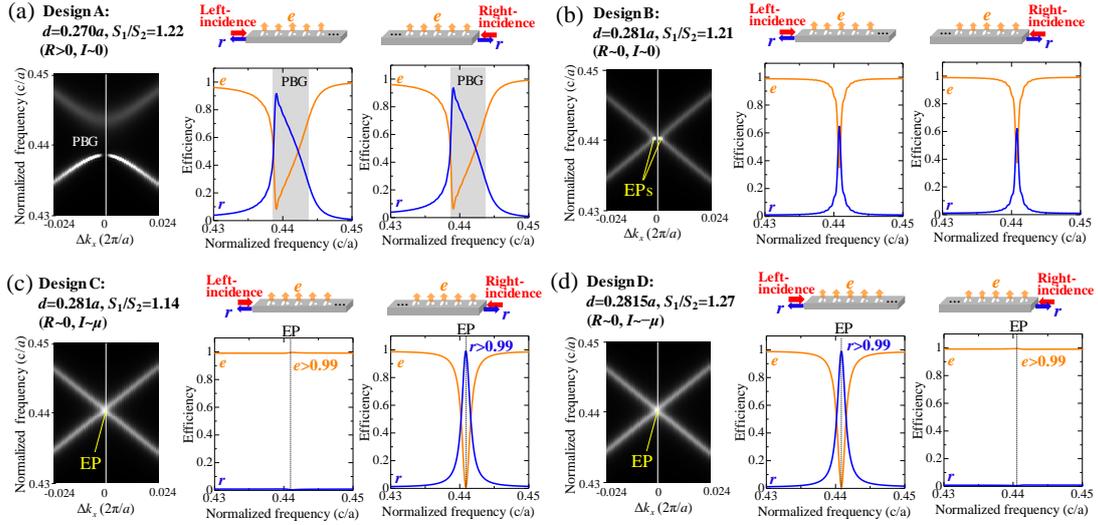

**FIG. 2.** Calculated photonic band structures and reflection/radiation spectra for left-incident and right-incident light for the designed double-lattice photonic-crystal waveguides: **(a)** Design A, **(b)** Design B, **(c)** Design C, and **(d)** Design D. Design C and Design D show nearly perfect radiation ($e$>0.99) for light incident from one side and nearly perfect reflection ($r$>0.99) for light incident from the other side at the frequency of the EP.

Then, we fabricate the designed unidirectional double-lattice photonic-crystal waveguides. A scanning microscope image of the fabricated double-lattice photonic crystal is shown in Fig. 3(a). Here, we employ a double-lattice photonic crystal composed of rectangular and square holes instead of elliptic and circular holes in Fig. 1(a) in order to simplify the fabrication process. The lattice constant $a$, the width $w$, the thickness $t$, and the total area of the two air holes ($S_1+S_2$) in the fabricated structures are $a$=750 nm, $w$=510 nm, $t$=220 nm, and $S_1+S_2=0.094a^2$, respectively. It should be noted that the deviation of these parameters and the hole shape from the ones used in our simulation do not inhibit the unidirectionality of the device itself, since the unidirectionality is mainly determined by the hole distance $d$ and the hole-size balance $S_1/S_2$. Here, we achieve unidirectionality by slightly increasing the hole distance ($d$) in order to preserve the linear dispersion with a single EP. We fabricate two devices with different double-lattice parameters: Device 1 ($d=0.278a$, $S_1/S_2=1.23$) and Device 2 ($d=0.298a$, $S_1/S_2=1.09$). We confirmed from 3D-FDTD simulations that the former corresponds to Design A ($R>0$, $I\sim0$) and the latter corresponds to Design C ($R\sim0$, $I\sim\mu$). Figure 3(b) shows an optical microscope image of the entire device, which includes two waveguides with oppositely oriented double-lattice photonic crystals to investigate the dependence on the direction of incidence.

Figure 3(c) shows a schematic of the setup for measuring radiation spectra from the double-lattice photonic crystals. Light from a wavelength-tunable laser is incident on the cleaved edge of the device using an objective lens. The radiation from the double-lattice photonic-crystal waveguide is collected by another objective lens and the radiative power is measured by using a power meter. Here, we use a mechanical chopper and a lock-in-amplifier for synchronous detection to improve the signal-to-noise ratio. It should be noted that each waveguide has a straight section without double-lattice photonic crystals ($L_{wg}=220$ µm) as shown in Fig. 3(b) so that we can collect the radiation from the double-lattice photonic crystals separately from the scattered light at the cleaved edge of the waveguide. Figures 3(d) and 3(e) show the measure radiation spectra of Devices 1 and 2 when the direction of incident light is reversed. For Device 1, the radiation power is minimum at around a wavelength of 1490 nm in both directions of incidence, which qualitatively agrees with the simulated result shown in Fig. 2(a). On the other hand, for Device 2 [Fig. 3(e)], the radiation power is almost independent of the wavelength for left-incident light, while it falls to nearly zero at around a wavelength of 1482 nm for right-incident light, indicating that unidirectional radiation (reflection) has been achieved.

For a more quantitative analyses, we focus on the fringe patterns appearing in the radiation spectra in Fig. 3(e). Figures 3(f) and 3(g) show enlarged views of the radiation spectra shown in Fig. 3(e). For right-incident light [Fig. 3(g)], a periodic interference pattern with a wavelength interval of $\Delta\lambda \sim 1.2$ nm and a maximum extinction ration of $\eta_{ext} > \sim 20$ is observed, which corresponds to the Fabry-Perot resonance inside the 220-µm-long straight waveguide between the cleaved edge of the device and the double-lattice photonic crystals. On the other hand, for left-incident light [Fig. 3(f)], the above-mentioned Fabry-Perot interference is not clearly observed ($\eta_{ext} < 2$), indicating that the reflectivity of the double-lattice photonic crystal is much smaller. The small, random fringes in Fig. 3(f) instead may be attributed to the scattering of light that cannot couple into the waveguide, but rather couples into the neighboring Si layer between the waveguides. Assuming that the power reflectivity at the cleaved edge and at the double-lattice photonic crystal are $r_0$ and $r$, respectively, the above-mentioned extinction ratio $\eta_{ext}$ of the Fabry-Perot interference can be derived as follows;

$$\eta_{ext} = \left( \frac{1+\sqrt{r_0 r}}{1-\sqrt{r_0 r}} \right)^2. \tag{7}$$

By substituting the measured extinction ratio $\eta_{ext}$ and the calculated reflectivity at the cleaved edge ($r_0 \sim 0.55$) into Eq. (7), the maximum reflectivity of the double-lattice photonic crystal in Device 2 can be estimated as $r < 0.06$ for left-incident light and $r > 0.73$ for right-incident light, which further indicates that unidirectional reflection (radiation) has been achieved in our device.

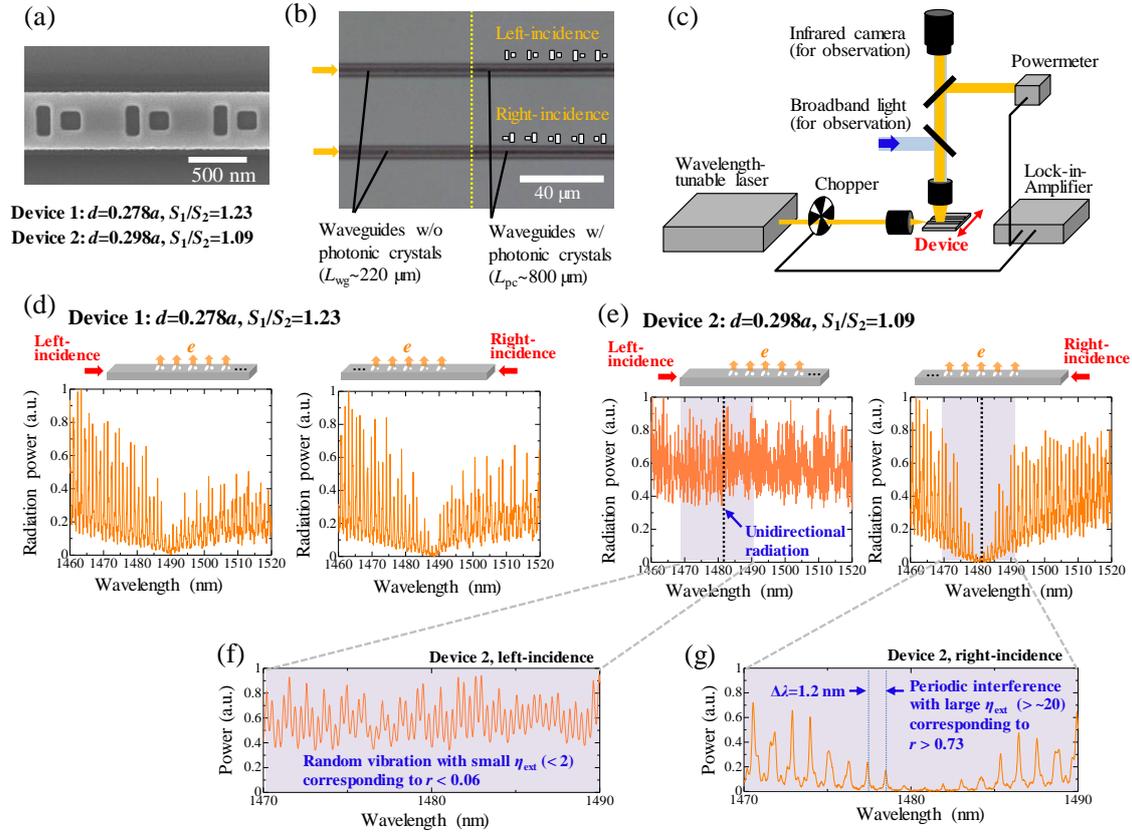

**FIG. 3. (a)** Scanning electron microscope image of a fabricated double-lattice photonic crystal waveguide. **(b)** Optical microscope image of the entire device. **(c)** Experimental setup for measuring the radiation spectra of the double-lattice photonic-crystal waveguides. **(d)(e)** Measured radiation spectra of Devices 1 ($R>0$, $I\sim 0$) and 2 ($R\sim 0$, $I\sim\mu$) when the incident direction of light is reversed. In Device 2 [Fig. 3(e)], unidirectional radiation is observed around a wavelength of 1482 nm. **(f)(g)** Enlarged view of the measure radiation spectra of Device 1 and 2 when the incident direction of light is reversed. $\eta_{ext}$ denotes the maximum extinction ratio of the Fabry-Perot resonance, which is less than 2 (corresponding to a reflectivity $r < 0.06$) for left-incident light [Fig. 3(f)], but more than ~20 (corresponding to a reflectivity $r > 0.73$) for right-incident light [Fig. 3(g)].

In conclusion, we have proposed and demonstrated unidirectional optical waveguides based on double-lattice photonic crystals, which show perfect radiation in the vertical direction when light is incident from one side and perfect reflection when light is incident from the opposite side. We have revealed that such unidirectional characteristics are based on a linear optical dispersion with a single EP, which can be realized by balancing the Hermitian and non-Hermitian couplings inside the double-lattice photonic crystals. Unlike previously demonstrated unidirectional waveguides, our proposed devices can achieve unidirectional radiation/reflection of one hundred percent of the incident light owing to the absence of material loss, and thus our devices enable various novel applications such as back-reflection-free vertical couplers for silicon photonics and highly efficient optical antennas for laser remote sensing. Our devices offer new degrees of freedom in the design of various nanophotonic devices harnessing non-Hermiticity, and they will greatly expand the utility of non-Hermitian devices in practical applications.

We thank X. Yin and J. Gelleta for useful discussions. This work was partially supported by a Grant-in-Aid for Scientific Research (22H04915, 20H02655) from the Japan Society for the Promotion of Science (JSPS).

**References**

[1] R. El-Ganainy, K. G. Makris, M. Khajavikhan, Z. H. Musslimani, S. Rotter, and D. N. Christodoulides, Non-Hermitian physics and PT symmetry, Nature Physics **14**, 11 (2018).

[2] M.-A. Miri and A. Alu, Exceptional points in optics and photonics, Science **363**, eaar7709 (2019).

[3] L. Feng, Z. J. Wong, R.-M. Ma, Y. Wang, X. Zhang, Single-mode laser by parity-time symmetry breaking, Science **346**, 972 (2014).

[4] H. Hodaei, M.-A. Miri, M. Heinrich, D. N. Christodoulides, and M. Khajavikhan, Parity-time-symmetric microring lasers, Science **346**, 975 (2014).

[5] Z. Lin, A. Pick, M. Loncar, and A. W. Rodriguez, Enhanced spontaneous emission at third-order dirac exceptional points in inverse-designed photonic crystals, Phys. Rev. Lett.

**117**, 107402 (2016).

[6] J. Doppler, A. A. Mailybaev, J. Bohm, U. Kuhl, A. Grischik, F. Libisch, T. J. Milburn, P. Rabl, N. Moiseyev, and S. Rotter, Dynamically encircling an exceptional point for asymmetric mode switching, Nature **537**, 76 (2016).

[7] W. Chen, S. K. Ozdemir, G. Zhao, J. Wiersig, and L. Yang, Exceptional points enhance sensing in an optical microcavity, Nature **548**, 192 (2017).

[8] T. Goldzak, A. A. Mailybaev, and N. Moiseyev, Light stops at exceptional points, Phys. Rev. Lett. **120**, 013901 (2018).

[9] Y.-H. Lai, Y.-K. Lu, M.-G. Suh, Z. Yuan, and K. Vahala, Observation of the exceptional-point-enhanced Sagnac effect, Nature **576**, 65 (2019).

[10] M. P. Hokmabadi, A. Schumer, D. N. Christodoulides, and M. Khajavikhan, Non-Hermitian ring laser gyroscopes with enhanced Sagnac sensitivity, Nature **576**, 70 (2019).

[11] Z. Lin, H. Ramezani, T. Eichelkraut, T. Kottos, H. Cao, and D. N. Christodoulides, Unidirectional invisibility induced by PT-symmetric periodic structures, Phys. Rev. Lett. **106**, 213901 (2011).

[12] L. Feng, Y.-L. Xu, W. S. Fegadolli, M.-H. Lu, J. E. B. Oliveira, V. R. Almeida, Y.-F. Chen, and A. Scherer, Experimental demonstration of a unidirectional reflectionless parity-time metamaterials at optical frequencies, Nat. Mater. **12**, 108 (2013).

[13] L. Feng, X. Zhu, S. Yang, H. Zhu, P. Zhang, X. Yin, Y. Wang, and X. Zhang, Demonstration of a large-scale optical exceptional point structure, Opt. Express **22**, 1760 (2014).

[14] Y. Huang, Y. Shen, C. Min, S. Fan, and G. Veronis, Unidirectional reflectionless light propagation at exceptional points, Nanophotonics **6**, 977 (2017).

[15] M. Yoshida, M. De Zoysa, K. Ishizaki, Y. Tanaka, M. Kawasaki, R. Hatsuda, B. Song, J. Gelleta, and S. Noda, Double-lattice photonic-crystal resonators enabling high-

brightness semiconductor lasers with symmetric narrow-divergence beams, Nature Mater. **18**, 121 (2019).

[16] T. Inoue, M. Yoshida, J. Gelleta, K. Izumi, K. Yoshida, K. Ishizaki, M. D. Zoysa, and S. Noda, General recipe to realize photonic-crystal surface-emitting lasers with 100-W-to-1-kW single-mode operation, Nat. Commun. **13**, 3262 (2022).

[17] B. Zhen, C. W. Hsu, Y. Igarashi, L. Lu, I. Kaminer, A. Pick, S.-L. Chua, J. D. Joannopoulos, and M. Soljacic, Spawning rings of exceptional points out of Dirac cones, Nature **525**, 354 (2015).

[18] A. Yulaev, S. Kim, Q. Li, D. A. Westly, B. J. Roxworthy, K. Srinivasan, and V. A. Aksyuk, Exceptional points in lossy media lead to deep polynomial wave penetration with spatially uniform power loss, Nat. Nanotech. **17**, 583 (2022).

[19] Y. Liang, C. Peng, K. Sakai, S. Iwahashi, and S. Noda, Three-dimensional coupled-wave model for square-lattice photonic crystal lasers with transverse electric polarization: A general approach, Phys. Rev. B **84**, 195119 (2011).

[20] Y. Liang, C. Peng, K. Sakai, S. Iwahashi, and S. Noda, Three-dimensional coupled-wave model for square-lattice photonic crystal lasers with transverse electric polarization: finite-size effects, Opt. Express **20**, 15945 (2012).

[21] T. Asano, M. Mochizuki, S. Noda, M. Okano, and M. Imada, A channel drop filter using a single defect in a 2-D photonic crystal slab-defect engineering with respect to polarization mode and ratio of emissions from upper and lower sides, J. Lightwave Technol. **21**, 1370 (2003).

[22] X. Yin, J. Jin, M. Soljacic, C. Peng, and B. Chen, Observation of topologically enabled unidirectional guided resonances, Nature **580**, 467 (2020).